# BINDING OF PHEOPHORBIDE-*a* METHYL ESTER TO NUCLEIC ACIDS OF DIFFERENT SECONDARY STRUCTURES: A SPECTROSCOPIC STUDY

O.A. Ryazanova [1], V.N. Zozulya [1], I.M. Voloshin [1], L.V. Dubey [2], I.Ya. Dubey [2]

[1]*Department of Molecular Biophysics, B. Verkin Institute for Low Temperature Physics and Engineering of NAS of Ukraine, 47 Nauky ave., 61103, Kharkiv,*
*e-mail: ryazanova@ilt.kharkov.ua*
[2]*Department of Synthetic Bioregulators, Institute of Molecular Biology and Genetics of NAS of Ukraine, 150 Zabolotnogo str., 03680 Kyiv, Ukraine*

Binding of neutral Pheophorbide-*a* methyl ester (MePheo-*a*) to various synthetic polynucleotides, double-stranded poly(A)·poly(U), poly(G)·poly(C) and four-stranded poly(G), as well as to calf thymus DNA, was studied using the methods of absorption and polarized fluorescent spectroscopy. Measurements were performed in aqueous buffered solutions (pH 6.9) of low ionic strength (2 мМ $Na^+$) in a wide range of molar phosphate-to-dye ratios (*P/D*). Absorption and fluorescent characteristics of complexes formed between the dye and biopolymers were determined. Binding of MePheo-*a* to four-stranded poly(G) is shown to be accompanied by the most significant spectral transformations: hypochromism of dye absorption, large bathochromic shift of Soret absorption band (~26 nm) and fluorescence band (~9 nm) maxima, 48-fold enhancement of the dye emission intensity. In contrast, its binding to the double-stranded polynucleotides and native DNA induces only small shifts of absorption and fluorescence bands, as well as no more than 4-fold rise of fluorescence intensity. Substantial spectral changes and high value of fluorescence polarization degree (0.26) observed upon binding of MePheo-*a* to quadruplex poly(G) allow us to suggest the intercalation of the dye chromophore between guanine tetrads. At the same time, small spectral changes and insignificant increase of MePheo-*a* fluorescence polarization degree (0.12) upon binding of the dye to double-stranded biopolymers point to another binding type. Incorporation of MePheo-*a* to a helix groove is supposed to occur, presumably in the dimeric form. Substantial enhancement of MePheo-*a* emission upon binding to four-stranded poly(G) allows us to propose this compound as a new fluorescent probe for G-quadruplex structures.

# I. INTRODUCTION

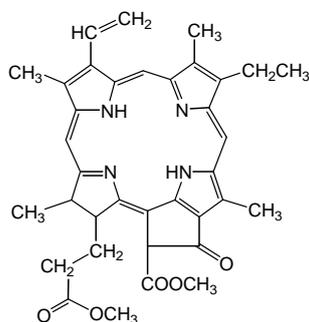

**Fig. 1.** Molecular structure of pheophorbide-*a* methyl ester (MePheo-*a*).

Increasing number of publication on interaction of the porphyrin derivatives with various biological targets that appeared during the last years can be explained by the wide application of these dyes in molecular biology and medicine, in particular, as photosensitizers in photodynamic therapy [1-3]. The chemical structure of Pheophorbide *a* (Pheo-*a*) is similar to pheophytin and chlorophyll-*a*, and the dye can be easily prepared from chlorophyll [4]. Due to high photosensitizing activity *in vitro* and *in vivo* [5-8], Pheo-*a* is widely used in photodynamic therapy of various tumors [2,9]. The photophysical properties of Pheo-*a* have been studied earlier [10, 11]. It was established that it selectively accumulates in cancer cells [12] and has a high extinction coefficient in the red region of the spectrum, where the permeability of living tissues to light increases significantly. The photodynamic activity of Pheo-*a* is mainly provided by monomeric dye molecules. It is significantly reduced by the dye dimerization [13] or aggregation. The photodynamic effect of Pheo-*a* is conditioned by both photoinduced oxidative processes caused by singlet molecular oxygen, $^1O_2$, which is actively generated by the dye irradiated with red light [10,12], and by processes caused by electron transfer between the DNA bases and the first excited singlet state of Pheo-*a*. It is established that Pheo-*a* causes the photo-fragmentation of DNA both in the presence and absence of oxygen [14]. However, Pheo-*a* possesses some disadvantages, namely, (i) its anionic character prevents the binding of the dye to negatively charged nucleic acids, and (ii) its low solubility in water complicates the delivery of the dye to the target. To improve the solubility of the dye and to simplify its binding to the nucleic acids, new cationic derivative of Pheo-*a* was synthesized (CatPheo-*a*), which has a side chain with a terminal trimethylammonium group [15].

Earlier from the study on the interaction of CatPheo-*a* with model polynucleotides of different secondary structures (double-stranded poly(A)·poly(U), poly(G)·poly(C) and four-stranded poly(G)) performed in aqueous buffered solutions containing 2 mM $Na^+$, two binding modes has been revealed. At low molar phosphate-to-dye ratios, *P/D*, the strong cooperative binding of the cationic dye to polynucleotide backbone was shown to be predominant (electrostatic interaction), which was accompanied by the stacking of neighboring chromophores [16]. This binding type was characterized by a strong quenching of CatPheo-*a* fluorescence and increase in its polarization degree [15]. With increasing *P/D* the dye associates disintegrated and another anticooperative intercalation type of binding became dominant. In the case of complexes containing double-stranded polynucleotides, its contribution in binding was insignificant, whereas for that containing a four-strand poly(G) it was large. For the CatPheo-*a* + poly(G) at high *P/D* values a 4.5-fold increase in the fluorescence intensity was observed as compared to its value for the free dye, and the degree of fluorescence polarization increased up to 0.26.

A study of the separate external electrostatic binding of Pheophorbide to nucleic acids was performed earlier with a model system consisting of CatPheo-*a* and single-stranded inorganic polyphosphate (PPS) [15]. It allowed to establish stoichiometry, cooperativity, and the binding constant, as well as to determine the spectroscopic characteristics of the complexes formed.

To investigate separately another type of Pheophorbide-*a* binding to nucleic acids dominating at high *P/D* values, a neutral derivative of this dye, its methyl ester (MePheo-*a*, Fig.1) was synthesized [17]. The aim of the present work was to study the interaction of MePheo-*a* with synthetic polynucleotides of different secondary structures (double-stranded poly(A)·poly(U), poly(G)·poly(C) and native DNA, and quadruplex formed by poly(G)) in aqueous buffer containing 2 mM $Na^+$ (pH 6.9) and to determine the absorption and fluorescence characteristics of the complexes formed.

## II. EXPERIMENTAL

Pheophorbide-*a* was purchased from Frontier Scientific (Logan, Utah, USA). Chromato-mass spectra (LC-MS) were obtained with Agilent 1100LC/MSD SL instrument (USA).

Polynucleotides poly(A)·poly(U), poly(G)·poly(C), poly(G) (Sigma Chemical Co.) were used without further purification. In all experiments, 2 mM Na-cacodylate buffer, pH 6.9, prepared from deionized distilled water was used as a solvent.

The concentration of polynucleotides was determined spectrophotometrically in aqueous solutions using the following molar extinction coefficients: $\varepsilon_{260} = 7140$ $M^{-1} cm^{-1}$ for poly(A)·poly(U), $\varepsilon_{260} = 7900$ $M^{-1} cm^{-1}$ for poly(G)·poly(C), $\varepsilon_{252} = 9900$ $M^{-1} cm^{-1}$ for poly(G). Since the solubility of MePheo-*a* in water is very low, its stock solution of known concentration in ethanol was initially prepared. To obtain an aqueous solution of the dye of a required concentration, the stock solution was diluted with a corresponding amout of aqueous buffer, so that the ethanol concentration in the samples did not exceed 10%.

Electron absorption spectra were obtained in a 0.5 and 1 cm quartz cell on a SPECORD UV-VIS spectrophotometer (Carl Zeiss, Jena, Germany). The intensity and polarization degree of the fluorescence were measured in a 0.5 cm quartz cell on a laboratory spectrofluorimeter based on a double DFS-12 monochromator (LOMO, Russia) by the photon counting method.

Fluorescence excitation was carried out using He-Ne laser with $\lambda_{exc}$ = 633 nm. The intensity of the laser radiation was weakened by means of a neutral color filter. The emission was recorded at the right angle to the exciting beam. The count of photons was made by accumulating pulses in 10 s for each experimental point. The experimental set-up and measurement procedure were described earlier [18]. The fluorescence polarization degree ($p$) was determined from the equation [19]:

$$p = \frac{I_{II} - I_{\perp}}{I_{II} + I_{\perp}} \quad (1)$$

where $I_{II}$ and $I_{\perp}$ are the components of polarized fluorescence that are parallel and perpendicular to the electric vector of the exciting light.

The binding of the dye to polynucleotides was studied by fluorescent titration method. A dye complex with a polynucleotide containing the same dye concentration but a higher polymer concentration was added to the solution of MePheo-$a$, which allowed to obtain the required ratio of the molar concentrations of polynucleotide phosphates and the dye, $P/D$. The concentration of MePheo-$a$ in all samples was $5\cdot10^{-6}$ M. During the experiment, the fluorescence intensity and polarization degree of the complex were recorded at the emission maximum of free MePheo-$a$. The measurements were carried out at room temperature (20-22 °C).

Pheophorbide-$a$ methyl ester was obtained by esterification of Pheo-$a$ in methanol under acid catalysis conditions as follows: Pheophorbide-$a$ (12 mg, 0.02 mmol), dried in vacuum over $P_2O_5$, was dissolved in 2 ml of anhydrous methanol, then 60 μl of concentrated sulfuric acid was added and the reaction mixture was allowed to stand overnight at room temperature. Chloroform (5 ml) was added, the mixture was washed with water (3 ml), saturated $NaHCO_3$ solution (2×3 ml) and again with water (3 ml). The organic layer was dried over anhydrous sodium sulfate. The product was applied onto a small column (1×5 cm) with silica gel

(Kieselgel 60, Merck). The column was washed with chloroform, the product was eluted with 0.5% methanol in chloroform. The appropriate fraction was evaporated in vacuum. The product was precipitated from chloroform into hexane, separated by centrifugation, washed with 0.5 ml of hexane and dried in vacuum to obtain 7 mg of almost black powder (yield 58%). LC-MS: *m/z* 607.5 [M+1]$^+$.

## III. RESULTS AND DISCUSSION

It is known that photophysical properties of Pheophorbide-*a* and its derivatives are mainly determined by delocalized π-electronic system of their macrocyclic chromophore [20]. In Figure 2 electronic absorption spectra of MePheo-*a* in ethanol and in aqueous solution containing 10 % of ethanol are presented. It can be seen from the figure that the absorption spectrum of the dye in ethanol consists of an intense B-band at 410 nm (the so-called Soret band), as well as four Q-bands whose electronic transition moments are known to lie in the plane of the molecule. The most intense of Q-bands is the long-wavelength one located at 666.5 nm. For the aqueous dye solution, 60% hypochromism of the absorption is observed being accompanied with a broadening of all bands, blue shift of the Soret band to 396.5 nm, and red shift of the Q bands (in particular, the long-wave band shifts to 676.5 nm). The shape of the MePheo-*a* absorption spectrum is typical for the porphyrin derivatives belonging to the subclass of chlorin [21].

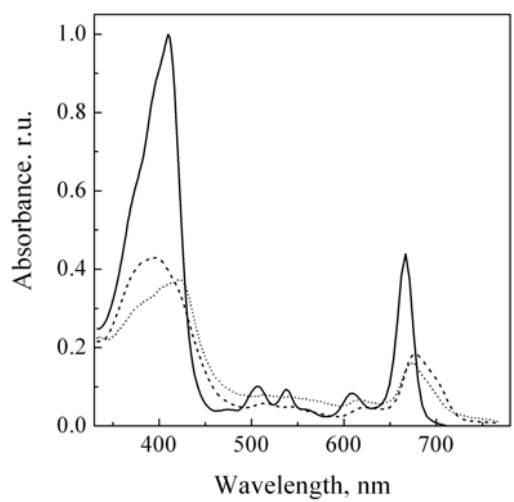 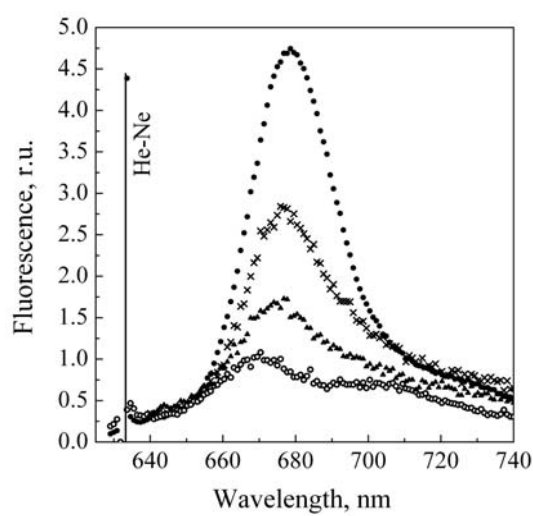

**Fig. 2**. Absorption spectra of free MePheo-*a* in ethanol (———) and water (------), and MePheo-*a* bound to poly(G) at $P/D = 458$ (······). Measurements were performed at 20 °C, the dye concentration was 10 μM. The spectra are normalized to the magnitude of the free dye absorption in ethanol at the wavelength of the Soret band maximum.

**Fig. 3.** Fluorescence spectra of MePheo-*a* in a free state (O) and bound to polynucleotides: poly(A)·poly(U) at $P/D = 3040$ (▲), poly(G)·poly(C) at $P/D = 408$ (✗), and poly(G) (intensity is reduced by 10 times) at $P/D = 458$ (●), $\lambda_{exc} = 633$ nm. Measurements were cariied out in Na-cacodylate buffer containing 10% of ethanol and 2 mM Na$^+$, $C_{MePheo} = 5$ μM.

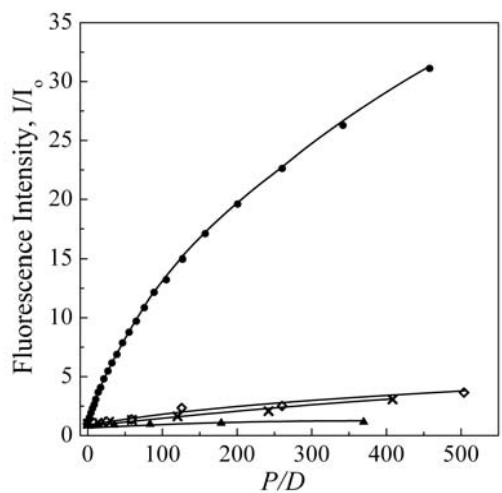 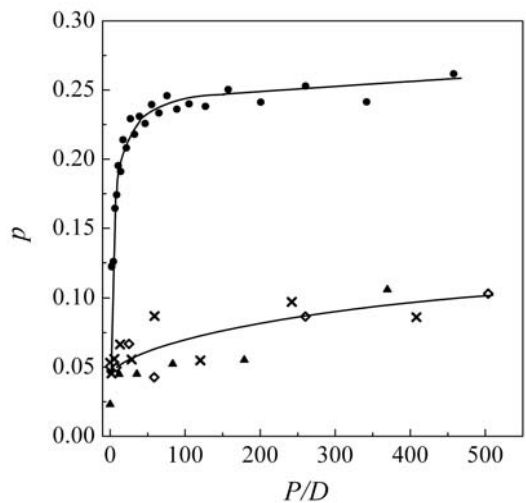

**Fig. 4.** Dependence of fluorescence intensity of MePheo-*a* at 669 nm on molar phosphate-to-dye ratio (*P/D*) when titrated with polynucleotides poly(A)·poly(U) (▲), poly(G)·poly(C) (✖), poly(G) (●) and calf thymus DNA (◇), $\lambda_{exc}$ = 633 nm. Measurements were performed in 2 mM sodium cacodylate buffer containing 10% of ethanol, $C_{MePheo}$ = 5 μM.

**Fig. 5.** Dependence of MePheo-*a* fluorescence polarisation degree at 669 nm on *P/D* molar ratio when titrated with polynucleotides poly(A)·poly(U) (▲), poly(G)·poly(C) (✖), poly(G) (●) and calf thymus DNA (◇), $\lambda_{exc}$ = 633 nm. Measurements were performed at the same condition as in Fig. 4.

The emission spectrum of MePheo-*a* in aqueous solutions (Fig. 3) represents a superposition of two bands, more intense of which has a maximum at 669 nm, and less intense broad band with a maximum at approximately 707 nm. The fluorescence polarization degree of the dye, *p*, measured at the emission band maximum is amounted to 0.01.

From the above data it follows that the fluorescence band of MePheo-*a* in aqueous solution (669 nm) is located in the shorter-wave region compared to its longwave absorption band (676.5 nm). The same mutual arrangement of these absorption and fluorescence bands was observed earlier for Pheo-*a* [22], CatPheo-*a* [15], and pyropheophorbide [23]. This fact seems, at first glance, paradoxical. However, it can be explained by the fact that in neutral aqueous solutions at the studied concentration these dyes exist predominantly in an dimeric form. The study of temperature dependence of the absorption and fluorescence intensities for CatPheo-*a* samples shows that the dissociation of these dimers occurs at temperatures above 70 °C [16]. Thus, it can be concluded that the longwave absorption band of MePheo-*a* in aqueous solutions corresponds to the absorption of the dye dimers, whereas the fluorescence band is due to the monomers. In ethanol solution the MePheo-*a* absorption is caused by the dye in a monomeric form.

Figures 4 and 5 show the curves of fluorescent titration of MePheo-*a* with double-stranded polynucleotides, poly(A)·poly(U), poly(G)·poly (C), native DNA, and four-stranded poly(G). The curves are plotted as the dependence of relative ntensity, $I/I_0$ (Fig. 4), and the polarization degree, $p$ (Fig. 5), of MePheo-*a* fluorescence on *P/D* ratio. Here $I_0$ is the fluorescence intensity of the free dye measured at the band maximum, 669 nm, and *I* is the fluorescence intensity of the dye-polynucleotide complex at the same wavelength. These data sugggest that an increase in the relative concentration of double-stranded polymers results in a gradual increase in both the dye fluorescence intensity and polarization degree. For example, at *P/D* = 400, the emission of MePheo-*a* + poly(A)·poly(U) complex is in 1.3 times more intense than that for the free dye, and with poly(G)·poly(C) and the native DNA it is 3 and 3.5 times higher, respectively. The fluorescence polarisation degree in all casese increases up to 0.1. Also, for all three complexes the maximum of the fluorescence band undergoes 7-8 nm red shift as compared with the maximum of free MePheo-*a* (Fig. 3).

The binding of MePheo-*a* to poly(G) is of special interest, since in aqueous solutions this polynucleotide can form four-stranded structure which can be a model of the specific structures of telomeric DNA, the so-called G-quadruplexes. A preliminary testing confirmed the formation of four-stranded poly(G) structure in the solution. For this purpose, the melting curve of the sample with the detection by absorption at 295 nm was obtained and analyzed, and the transition induced by unfolding of the four-strand structure [24] was clearly manifested.

From the fluorescent titration curves plotted for the MePheo-*a* + poly(G) system (Fig. 4) it is seen that an increase in the relative polymer content results in a sharp enhancement of the MePheo-*a* fluorescence. So, at *P/D* = 400 its intensity measured at 669 nm is 30 times higher than that for the free dye (*P/D* = 0) at the same wavelength, while the ratio between the fluorescence intensities measured in the band maximum for the MePheo-*a* + poly(G) complex (at 678.5 nm) and free MePheo-*a* (at 669 nm) reaches 48. Also an increase in *P/D* induces 26 nm bathochromic shift of the dye Soret band and 3 nm hypsochromic shift of the longwave Q-band in the absorption spectrum (Fig. 2), as well as 9 nm bathochromic shift of the fluorescence band maximum (up to 678.5 nm). The observed spectral changes can be caused by the chromophore withdrawal from water environment as a result of its intercalation between the guanine bases. This hypothesis is supported by the high value of the fluorescence polarization degree, which at high *P/D* value reaches a constant level of 0.26 (Fig. 5).

Comparison of the fluorescence titration curves for MePheo-*a* + poly (G) with those for CatPheo-*a* + polya(G) [16] shows that fluorescence enhancement observed for the neutral porphyrin at high *P/D* values is substantially stronger than that for the cationic CatPheo-*a* derivative. So at *P/D* = 400 the $I/I_0$ ratio determined at 669 nm (the wavelength corresponding to the emission band maximum of the free dye) is amounted to 30 and 4.5, respectively. This difference can be explained by stronger dimerisation of the neutral MePheo-*a* in water solution at *P/D = 0*, as well as by the significant contribution of external electrostatic binding of positively charged CatPheo-*a* to the negatively charged phosphate backbone of the biopolymers, which is accompanied by the self-stacking of the dye chromophores and results in quenching of their fluorescence. Although the largest contribution of electrostatic binding occurs at low *P/D* ratios, even at a high relative concentration of the polymer it remains sufficient.

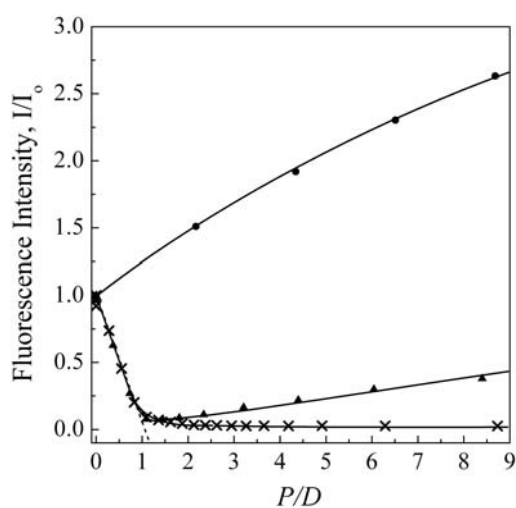

**Fig. 6.** Comparison of the fluorescence intensity measured at a maximum of the free dyes bands on *P/D* molar ratio for MePheo-*a* · poly(G) (●), CatPheo-*a* · poly(G) (▲) and CatPheo-*a*· PPS (✕), $\lambda_{exc}$ = 633 nm.
Measurements were performed in 2 mM Na-cacodylate buffer. MePheo-*a* solution contained 10% of ethanol. Dye concentrations were $5 \cdot 10^{-6}$ M for the samples with poly(G) and $1.3 \cdot 10^{-5}$ M for that with PPS.

Fig. 6 shows a comparison of the fluorescent titration curves at low *P/D* values for MePheo-*a* + poly(G) with those reported earlier for CatPheo-*a* + poly(G) [16] and CatPheo-*a* + PPS (single strand inorganic polyphosphate) [15]. It can be seen that for the last two systems titration curves are biphasic. For the CatPheo-*a* + PPS complex in the *P/D* range from 0 to 1, a sharp 40-fold quenching of CatPheo-*a* fluorescence was observed that can be explained by self-stacking of the neighbored CatPheo-*a* chromophores electrostatically boundg to polyanion exterior. At *P/D* > 1 the increase in *P/D* results in a slight rise of the fluorescence intensity, reaching at *P/D* = 10,000 only 10% from its initial value [15]. At the same time for the CatPheo-*a* + poly(G) complex, which titration curve coincides with that for CatPheo-*a* + PPS in low *P/D* range of 0–1, the increase of relative polymer concentration at *P/D* > 1 leads to noticeable enhancement of CatPheo-a emission. We suppose that it is conditioned by another binding type, presumably, by intercalation of the dye chromophore between guanine tetrads. A completely different picture was observed for the binding of

MePheo-*a* to poly(G) which is characterised by monophasic titration curve (Fig. 6). Monotonic enhansement of the dye emission starts right after $P/D = 0$. We suppose that for neutral MePheo-*a* compound only one binding type, intercalation, is realized. It should be noted that such emission enhancement observed for Pheophorbide-*a* derivatives bound to poly(G) is of particular interest, since the emission of most fluorescent dyes is known to be effectively quenched by guanine bases [25, 26].

Comparison of fluorescence titration curves for Mepheo-*a* with double- and four-stranded polynucleotides (Figures 4 and 5) shows higher binding affinnity in the case of quadruplex polymer. So, for Mepheo-*a* bound to poly(G) at high *P/D* ratios the fluorescence intensity measured at 669 nm (that corresponds to maximum of Mepheo-*a* emission in a free state) is 30 times higher than that for the free dye (at $P/D = 0$), whereas in the case of poly(A)·poly(U) this factor is only 1.7, and for poly(G)·poly(C) and native DNA it lies in the range of 3 – 3.7. The increase in *P/D* induces also the rise of fluorescence polarization degree up to 0.26 in the case of poly(G), and only to 0.1 when the dye is bound to the double-stranded polymers (Figure 5). Comparison of the absorption spectra (Figure 2) shows that substantially more pronounced changes are observed upon binding of the dye to quadruplex polymer. While binding of MePheo-*a* to double-stranded polynucleotides results in 7 nm red shift of the Soret band, in the case of poly(G) the shift is 26 nm. At the same time, for both systems the position of the Q-bands remains practically unchanged. The literature data available on the binding of mono-, di-, tetra- and hexa-cationic derivatives of pyropheophorbide-*a* [27], which structure is very similar to that of Pheophorbide-*a*, to double-stranded DNA performed using gel electrophoresis and absorption titration techniques show that intercalation of the dye into a polymer double helix was observed only for tetra- and hexacationic derivatives. However, the sensitivity of the gel-electrophoresis and absorption spectroscopy methods is significantly lower than that of fluorescence spectroscopy. Therefore, weak intercalation binding could not be detected by these two methods. Since the constant of binding of neutral Pheo-*a* methyl ester to negatively charged biopolymers is obviously significantly lower than that for the cationic Pheophorbide-*a* [15, 16] and pyropheophorbide-*a* [27] derivatives, we assume that in the case of double-helix polynucleotides and DNA, the dye can incorporates into their groove, presumably in the form of dimers, that results in the observed small increase in the MePheo-*a* fluorescence intensity and polarization degree.

The spectroscopic properties of MePheo-a and its complexes with synthetic polynucleotides are given in Table 1.

**Table 1.** Spectroscopic properties of MePheo-*a* in a free state and bound to polynucleotides.

| Sample | Absorption | | Fluorescence | | |
|---|---|---|---|---|---|
| | B-band: $\lambda_{max}$, nm | Longwave Q-band: $\lambda_{max}$, nm | $\lambda_{max}$, nm | $p$ | $I/I_0$ at 669 nm |
| MePheo-*a* in ethanol | 410 | 666.5 | 673.5 | 0.01 | 150 |
| MePheo-*a* in ABS* | 396.5 | 676.5 | 669 | 0.02 – 0.05 | 1 |
| MePheo-*a* + poly(A)·poly(U) (*P/D* = 400) in ABS* | 403 | 678 | 677 | 0.1 | 1.3 |
| MePheo-*a* + poly(G)·poly(C) (*P/D* = 400) in ABS* | 403.5 | 676 | 677 | 0.1 | 3 |
| MePheo-*a* + poly(G) (*P/D* = 400) in ABS* | 422.7 | 673 | 678.5 | 0.26 | 30 |

* ABS – aqueous buffered solution: 2 mM Na-cacodylate buffer (pH 6.9)

A significant (up to 48-fold) enhancement of MePheo-*a* emission upon its intercalation between guanine tetrads of poly(G) is a very important feature of this compound. This effect allows us to propose this dye as a new fluorescent probe for the quadruplex structures of nucleic acids.

## IV. CONCLUSIONS

Study on binding of neutral Pheophorbide-*a* derivative, Pheo-*a* methyl ester, to synthetic polynucleotides of various base compositions and secondary structures suggests that the dye interacts with poly(G) by means of intercalation of its chromophores between neighboring guanine tetrads of the polymer. On the other hand, its binding to double-stranded poly(A)·poly(U), poly(G)·poly(C) polynucleotides and native DNA is realized by embedding the dye into the groove of the double helix, presumably in the dimeric form.

Since binding of MePheo-*a* to four-stranded poly(G) is accompanied by a significant enhancement of the dye emission, MePheo-*a* can serve as an efficient fluorescent light-up probe for the recognition and detection of G-quadruplex structures of nucleic acids.

## ACKNOWLEDGEMENT

This work was in part supported by the NAS of Ukraine program "Molecular and Cellular Biotechnologies for Medicine, Industry and Agriculture" (grant 43/18).